\documentclass[aps,reprint,prl,floatfix]{revtex4-1}
\usepackage{amssymb}
\usepackage{graphicx}
\usepackage{dcolumn}
\usepackage{bm}
\usepackage{hyperref}

\newcommand{\ux}{u_{\mathrm{L}}}
\newcommand{\uy}{u_{\mathrm{T}}}
\newcommand{\fx}{f_{\mathrm{L}}}
\newcommand{\fy}{f_{\mathrm{T}}}

\begin{document}

\title{Power exponential velocity distributions in disordered porous media}

\author{Maciej Matyka} \email{maciej.matyka@ift.uni.wroc.pl}
\author{Jaros{\l}aw Go{\l}embiewski}
\author{Zbigniew Koza}

\affiliation{Faculty of Physics and Astronomy, University of Wroc{\l}aw,
 50-204 Wroc{\l}aw, Poland}

\begin{abstract}

Velocity distribution functions link the micro- and macro-level
theories of fluid flow through  porous media.
Here we study them for the fluid absolute velocity and its longitudinal
and lateral components relative to the macroscopic flow direction in a model of a
random porous medium.
We claim that all distributions follow the power exponential law
controlled by an exponent $\gamma$ and a shift parameter $u_0$ and
examine how these parameters depend on the porosity. We find that
$\gamma$ has a universal value $1/2$ at the percolation
threshold and grows with the porosity, but never exceeds~2.

\end{abstract}

\maketitle

The physics of viscous flows through porous media is important in such diverse areas
of technology as oil recovery, energy storage, and tumor treatment \cite{Olivieira11,Li15,Penta15}.
Such flows, however, are notorious for their complexity stemming both from randomness
of the medium and complicated interactions of different fluid particles.
Macroscopic parameters characterizing fluid transport in porous media,
like  permeability
(the ability of a porous system to transmit fluids)
depend on a multitude of geometry-related parameters such as
porosity, granule (or fracture) shape and size distribution, and specific surface
area. This dependency, however, is nonuniversal and to a large extent known only through phenomenology or approximate theories.

The complete information about the flow of an incompressible fluid in a particular porous sample
is contained in the velocity field.
While this quantity can be studied both experimentally
\cite{Kutsovsky96,Mansfield1996a,Datta13,Morad09} and
numerically \cite{Andrade97,Andrade99},
it is sample-dependent. 
Therefore, to get a better insight into the connection between the macroscopic properties of
the flow  and the 
irregular structure of the medium, one needs
mathematical tools that take into account randomness of the porous matrix and filter out irrelevant,
sample-dependent information contained in the full velocity field.
One such tool is the velocity distribution function (vdf)
\cite{Kutsovsky96,Mansfield1996a,Datta13}, which is the probability density function
of the fluid velocity magnitude $u$ or its  longitudinal ($\ux$) or transverse ($\uy$) components.
We will use a convenience notation $f$, $\fx$, and $\fy$ to denote
the vdfs corresponding to $u$, $\ux$, and $\uy$, respectively,
and $\fx^+$ and $\fx^-$ to denote $\fx$ restricted to $\mathbb{R}_{> 0}$ and $\mathbb{R}_{< 0}$, respectively.

Unlike the famous Maxwell-Boltzmann distribution for the ideal gas,
vdfs for a fluid flow reflect the structure of the medium rather than the effect
of the inter-particle collisions. Despite this difference, the vdfs are also closely related
to important macroscopic parameters. 
For example, $f$ and $\fx$ immediately imply the value of the hydraulic tortuosity ($\tau$) \cite{Duda11},
a quantity that measures the mean elongation of fluid paths in a porous medium 
\begin{equation}
 \label{eq:tort}
   \tau \equiv
   \frac{\langle u \rangle}{{\langle \ux \rangle}} =
   \frac{\int_V f(u) u \, \mathrm{d} u}{\int_V \fx(\ux) \ux  \, \mathrm{d}\ux} ,
\end{equation}
where the integrals are taken over the volume $V$ of the porous sample.
Similarly, for flows obeying Darcy's law \cite{Bear72}, e.g. groundwater flows,
the permeability ($\kappa$) can be related to the mean fluid velocity
along the macroscopic flow direction
\begin{equation}
 \label{eq:Darcy}
   \kappa = \varphi\mu\frac{\langle \ux \rangle }{|\nabla P|} ,
\end{equation}
where  $\varphi$ is the porosity of the medium, $\mu$ is the dynamic viscosity of the fluid,
and $ \nabla P$ is the pressure gradient.
Thus, the vdfs can serve as a link between two macroscopic parameters, $\kappa$ and $\tau$.

Several reports on $f$, $\fx$, and $\fy$ for various porous systems
at low Reynolds number ($\mathrm{Re} \ll 1$) are already available.
Physically, for arguments much smaller than  $\langle u \rangle$ their form is dominated
by contributions from stagnant zones (dead-end pores and the volumes
in the proximity of the fluid-solid boundary) \cite{Andrade97},
whereas for arguments $\gtrsim \langle u \rangle$ their form reflects the properties
of the conducting backbone, or the fluid paths carrying most of the fluid transport.
For this reason the vdf is usually investigated in two physically distinct regimes:
small ($u \lesssim \langle u \rangle$) and high ($u \gtrsim \langle u \rangle$) fluid velocities.
In the former case, the local
fluid kinetic energy at percolation follows a power law  \cite{Andrade97},
which implies a similar, power-law behavior for $f(u)$, $u \ll \langle u \rangle$.
Far from percolation, however, the form of $f(u)$ for small $u$  depends on the porous matrix structure \cite{Bijeljic13}
and appears to be nonuniversal, therefore it will not be considered here.

In contrast to the case of small velocities, the available results suggest the existence of some universality
in the form of the vdfs for $u \gtrsim \langle u \rangle$.
The findings of different research groups, however, appear to be inconsistent with each other.
On the one hand, several theoretical \cite{Mansfield1996a}, experimental \cite{Mansfield1996a}
and numerical \cite{Bijeljic13,Araujo06} results suggest that $f(u)$ can be approximated
by a Gaussian with the maximum shifted towards the mean fluid velocity.
On the other hand, however, several teams reported nearly exponential vdfs with the maximum located at 0.
This includes an experimental study on $f$, $\fx^+$, and $\fy$ \cite{Datta13}, as
well as experiments  \cite{Kutsovsky96,Lebon96} and numerical simulations \cite{Lebon96,Maier99} for $\fx^+$.
Moreover, a qualitative transition from an exponential to a Gaussian form of $f$ was found
for various sphere packings \cite{Maier96}; however, in each case  $f$ peaked at $u=0$.
Finally, Siena et~al.~\cite{Siena14} suggested that  $\fx^+$ follows  the stretched exponential function
\begin{equation}
 \label{eq:Siena}
  \fx^+ (\ux/\langle \ux \rangle) \propto
     \left(
       \ux/\langle \ux \rangle
     \right)^{\gamma-1}
      \exp\left[-\beta\left(  \ux/\langle \ux \rangle \right)^\gamma \right],
\end{equation}
where $\beta, \gamma$ are model parameters.

Although Eq.~(\ref{eq:Siena}) encompasses both the exponential and Gaussian distributions,
it is not applicable to the systems with the distribution maximum shifted from 0.
As a consequence, it predicts that the values of $\gamma$ can be much larger than~2 \cite{Siena14},
a result not corroborated by any other research.

To reconcile this difficulty, we conjecture that for $u \gtrsim \langle u \rangle$ 
the velocity distribution functions follow the exponential power distribution
\begin{equation}
\label{eq:fit-formula} 
  f(u) = a\exp\left[-\left(\frac{u-u_0}{u_\mathrm{w}}\right)^\gamma\right]
\end{equation}
(and similar formulas for $\fx^+$, $\fx^-$, and $\fy$),
where $a >0 $ is the normalizing factor, $u_0 \ge 0$ determines the location of the distribution peak,
$u_\mathrm{w} >0 $ denotes the scale factor corresponding to the distribution width,
and $0 < \gamma \le 2$ is the shape factor.
This is the simplest distribution that
generalizes  both the normal  ($\gamma=2$) and Laplace ($\gamma=1$) distributions and allows for the shift
of the distribution maximum from 0. In particular, in contrast to (\ref{eq:Siena}),
the prefactor to the exponential function in (\ref{eq:fit-formula}) does not depend on  $\mathbf{u}$.
We also postulate that for $f$ and $\fx^+$ there exists a threshold value of the porosity, $\varphi^*$, such that
\begin{equation}
 \label{eq:bounds}
  \begin{array}{rcl}
    u_0 = 0    &  \mbox{ for }  & \varphi_\mathrm{c} \le \varphi < \varphi^*,\\
    \gamma = 2 &  \mbox{ for }  & \varphi^* \le \varphi < 1,
  \end{array}
\end{equation}
which reduces, by 1, the number of unknown parameters in (\ref{eq:fit-formula}) for any given $\varphi$.
This number can be reduced to 2 by noticing that each vdf is normalized to 1.

To verify Eq.~(\ref{eq:fit-formula}),
we examined numerically an effectively  two-dimensional model
of fibrous materials with  the porous matrix
built of identical,  freely overlapping objects randomly deposited on a regular lattice of size $L$ \cite{Tomadakis93}.
We  considered two obstacle shapes, disks and squares, both with the hydraulic diameter $a=8$ lattice units (l.u.).
The fluid was assumed to be incompressible and Newtonian, driven by a bulk force (gravity)
small enough to ensure the creeping flow (Re $\ll 1$).

The basic numerical method used to solve the problem was the Palabos (www.palabos.org) implementation of
the lattice Boltzmann method (LBM)  with the Bhatnagar-Gross-Krook approximation for collisions
and the numerical viscosity $\nu = 1/6$
\cite{Succi01,Huber13,Pan06}.
While the LBM is often used for solving flows in porous media, its accuracy decreases when
the channels in the porous matrix are too narrow.
To verify whether this effect is significant in the model of overlapping objects,
we also solved it with the finite difference (FD) method. 
In this case we used only square obstacles arranged so that the minimum channel width was 4~l.u.\ \cite{Succi01}.
In both cases we used the periodic boundary conditions along the macroscopic fluid flow direction.
As for the transverse direction, we applied the no-slip boundary conditions in the LBM and periodic ones for the FD.
To minimize the finite-size effects, the lattice size, $L=1000$~l.u.\ (LBM) and 2000~l.u.\ (FD)
was chosen to ensure that  $L/a>100$ \cite{Koza09} and the results were compared
with those obtained for the system of size $L/2$.
The simulation results were averaged over 20 independent porous samples
for porosities $\varphi=0.99, 0.95, 0.9,\ldots$ down to the proximity of the percolation threshold
$p_\mathrm{c}\approx 0.4968$ for the squares \cite{Koza14} and $\approx 0.40$ for the disks.
The fluid velocity was measured at the underlying lattice nodes and binned to create histograms.

\begin{figure*}
  \includegraphics[width=1.7\columnwidth]{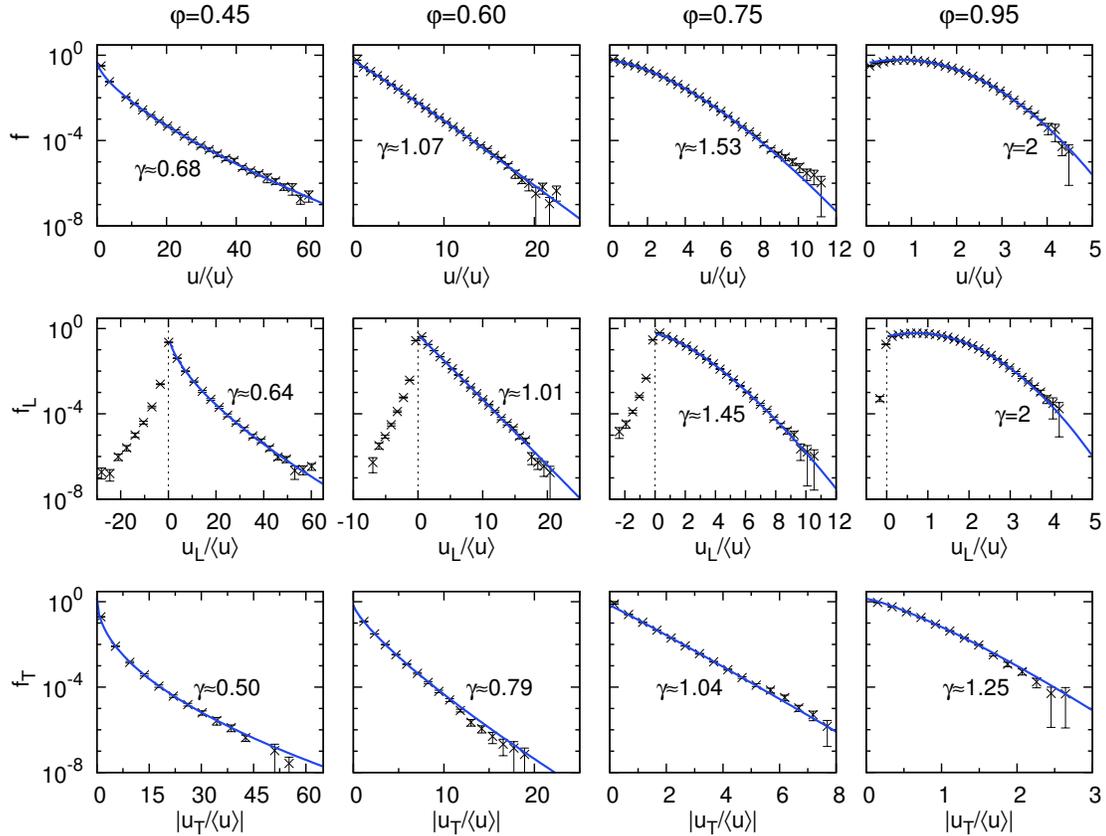}
\caption{\label{fig:4x3panel}
  (Color online) The probability distributions  $f$ (top row), $\fx$ (middle),
  and $\fy$ (bottom) for selected porosities $\varphi=0.45,0.60, 0.75, 0.95$
  (in columns, from left to right) obtained for overlapping disks.
  Solid lines show  fits to Eq.~(\protect\ref{eq:fit-formula}).}
\end{figure*}

\begin{figure}
  \includegraphics[width=0.9\columnwidth]{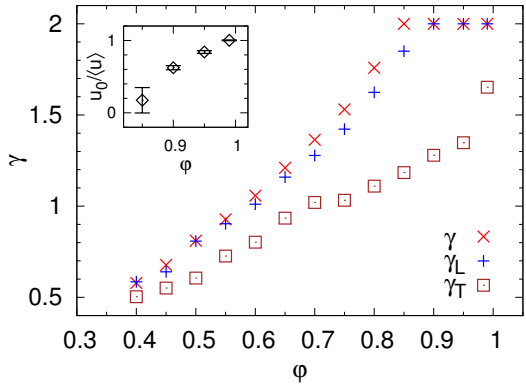}
\caption{\label{fig:exponents}
 (Color online) Exponent  $\gamma$ 
 for $f$ ($\times$), $\fx^+$ ($+$), and $\fy$ ($\square$) as a function of  porosity for overlapping disks.
 The symbol size roughly corresponds to the uncertainty of the results.
 Inset: $u_0/\langle u \rangle$ as a function of porosity for $f$ and $\varphi > \varphi^*\approx 0.85$.
 }
\end{figure}

\begin{figure}
\includegraphics[width=0.9\columnwidth]{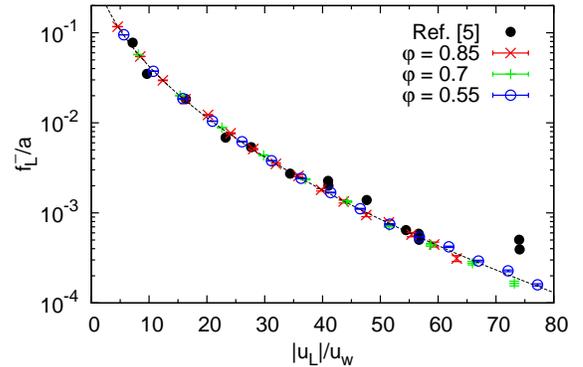}
\caption{
   \label{fig:ux_minus}
 (Color online) Scaling of $\fx^-$ according to Eq.~(\protect\ref{eq:fit-formula})
 with $u_0=0$ and $\gamma$ fixed at $0.5$.
 Open symbols show the results for overlapping squares
 at porosity $\varphi = 0.55,0.7$,  $0.85$; filled circles represent the experimental data
 of Ref.~\protect\cite{Datta13}.
 The dashed line is the scaling function $g(x) = \exp(-\sqrt{x})$.
}
\end{figure}

Representative results obtained for overlapping disks using the LBM are shown in
Fig.~\ref{fig:4x3panel}.
All velocities in this figure are normalized by $\langle u \rangle$,
and hence effectively dimensionless.
All data for $f, \fx^+, \fy$,  including those not shown, can be fitted well
to Eq.~(\ref{eq:fit-formula}) constrained by Eq.~(\ref{eq:bounds}).
As the porosity is increased from $\varphi_\mathrm{c}$ towards 1, a semilog plot of $f$ changes its shape
from convex (subexponential), through linear (exponential), concave (superexponential),
parabolic (Gaussian centered at 0) and shifted parabolic (Gaussian shifted towards $\langle u \rangle$).
The form of $\fx^+$ closely follows that of $f$, but $\fx^-$ vanishes faster than $\fx^+$,
especially far from $\varphi_\mathrm{c}$.
The  exponent $\gamma$ corresponding to  $\fy$ also turns out to be  $\varphi$-dependent,
though  its value never reaches 2 (Fig.~\ref{fig:exponents}).
This extends the experimental findings of Ref.~\cite{Datta13}, where $\gamma\approx 1$ was reported for $\fy$
at a fixed $\varphi$ that was chosen far from both $\varphi_\mathrm{c}$ and 1.
Note, however, that the current simulations cannot be used to reliably estimate $\gamma$ for $\fy$
in the limit of $\varphi\to 1$, as in this limit the number of obstacles becomes very small and hence
a large value of $L$ is required to avoid finite-size effects.
It is thus possible that in this limit $\gamma$ tends to 2.
The threshold value $\varphi^*\approx0.85$ for $f$ is close to its counterpart  $\approx 0.87$ for $\fx^+$;
the accuracy of our simulations was insufficient to tell if they are actually different from each other.
As expected, $u_0$  turns out  a continuous function of $\varphi$, growing from 0
for  $\varphi \le \varphi^*$ to $\langle u \rangle$ for $\varphi=1$ (Fig.~\ref{fig:exponents}, inset).
Similar results were obtained for overlapping squares (data not shown).
As for $\fx^-$, which controls the probability distribution of negative values of $u_\mathrm{L}$,
we found that its tail is well described by exponential power distribution
with $\gamma = 0.5$ for all $\varphi$ (Fig.~\ref{fig:ux_minus}).
This conclusion was drawn from  simulations 
for $\varphi \le 0.85$, as for
higher porosities the decay of $\fx^-$ is so rapid, cf.~Fig.~\ref{fig:4x3panel},
that no reliable fitting of the data is possible.
The value of $\gamma=0.5$ is also consistent with the experimental results obtained recently in Ref.~\cite{Datta13}
(Fig.~\ref{fig:ux_minus}) and our simulation data for overlapping disks (data not shown).
The contribution of $\fx^-$ to the transport is negligible compared  to that of $\fx^+$ only for relatively high porosities.
Consequently, simplified theories that assume $\fx^- \equiv 0$ \cite{Fleurant2001} are invalid close to $\varphi_\mathrm{c}$.

Thus, the following general picture emerges. At the percolation threshold
all four velocity distribution functions, $f, \fx^+, \fx^-$, and $\fy$
decay in accordance with the power exponential distribution
(\ref{eq:fit-formula}) with $u_0=0$ and $\gamma=1/2$. As the porosity is increased, $\gamma$
remains constant for $\fx^-$, but increases for $f, \fx^+$, and $\fy$.
At some threshold porosity $\varphi^*$ the exponent $\gamma$ determined
for $f$ or $\fx^+$ reaches the maximum value of 2.
As the porosity is increased above $\varphi^*$, $\gamma$ stays fixed at 2 whereas parameter $u_0$ becomes $\varphi$-dependent and grows from 0 to $\langle u \rangle$ as $\varphi$ approaches  1.
Therefore, a vdf in a random porous medium is either subexponential ($\gamma < 1$), exponential ($\gamma = 1$),
superexponential ($2 > \gamma > 1$), or normal ($\gamma = 2$).
If $\gamma=2$, the corresponding vdf depends on the porosity through the shift parameter $u_0$.

The significance of Eq.~(\ref{eq:fit-formula}) is related to several factors.
First, it describes the statistical properties of the microscopic velocity field for the  velocities $u\gtrsim \langle u \rangle$
that have a major contribution to the advective transport.
Second, it is a relatively simple formula with only two unknown parameters at any given porosity.
Third, the dependence of at least one of these parameters, $\gamma$, on the porosity also appears to be
fairly simple: in the model studied here it could be roughly approximated with two straight line segments (Fig.~\ref{fig:exponents}).
Given this simplicity, Eq.~(\ref{eq:fit-formula}) might be used to link porosity with vdf-dependent macrospcopic parameters like permeability or tortuosity.
It should also be  useful in studies on several open issues, like the microscopic foundations of permeability and
hydrodynamic dispersion (longitudinal and transverse) of passive solutes \cite{Sahimi93,Hunt2009,Icardi14},
the physical relevance of the tortuosity \cite{Ghanbarian2013}, and properties  of the conducting backbone
\cite{Hunt2009}, all with immediate practical applications.

However, some important questions remain open. For example, to what extent our hypothesis
remains valid for porous matrices with a complex, highly correlated structure~\cite{Rong13}?
Another problem is whether the value of $\gamma=2$ for $f$ and $\fx^+$ actually implies
the absence of long-range correlations in the velocity field above $\varphi^*$ \cite{Datta13}
or perhaps these correlations do remain and require that the power exponential
distribution  be supplemented with some less significant terms?

In summary, we propose that  a general form of the velocity distribution functions in
disordered porous media is given by a power exponential distribution with the shape factor $\gamma$
and location parameter $u_0$ such that  $1/2 \le \gamma \le 2$ and either $u_0=0$ or $\gamma=2$.
Moreover, $\gamma$ has a universal, porosity-independent value 1/2
both at the percolation threshold and for the negative part of the velocity component parallel to the macroscopic fluid flow direction.
Our findings resolve several apparently conflicting reports
on the velocity distribution functions
and open a new perspective on the formulation of a statistical theory of transport in porous media.

\bibliography{tort}

\begin{thebibliography}{29}%
\makeatletter
\providecommand \@ifxundefined [1]{%
 \@ifx{#1\undefined}
}%
\providecommand \@ifnum [1]{%
 \ifnum #1\expandafter \@firstoftwo
 \else \expandafter \@secondoftwo
 \fi
}%
\providecommand \@ifx [1]{%
 \ifx #1\expandafter \@firstoftwo
 \else \expandafter \@secondoftwo
 \fi
}%
\providecommand \natexlab [1]{#1}%
\providecommand \enquote  [1]{``#1''}%
\providecommand \bibnamefont  [1]{#1}%
\providecommand \bibfnamefont [1]{#1}%
\providecommand \citenamefont [1]{#1}%
\providecommand \href@noop [0]{\@secondoftwo}%
\providecommand \href [0]{\begingroup \@sanitize@url \@href}%
\providecommand \@href[1]{\@@startlink{#1}\@@href}%
\providecommand \@@href[1]{\endgroup#1\@@endlink}%
\providecommand \@sanitize@url [0]{\catcode `\\12\catcode `\$12\catcode
  `\&12\catcode `\#12\catcode `\^12\catcode `\_12\catcode `\%12\relax}%
\providecommand \@@startlink[1]{}%
\providecommand \@@endlink[0]{}%
\providecommand \url  [0]{\begingroup\@sanitize@url \@url }%
\providecommand \@url [1]{\endgroup\@href {#1}{\urlprefix }}%
\providecommand \urlprefix  [0]{URL }%
\providecommand \Eprint [0]{\href }%
\providecommand \doibase [0]{http://dx.doi.org/}%
\providecommand \selectlanguage [0]{\@gobble}%
\providecommand \bibinfo  [0]{\@secondoftwo}%
\providecommand \bibfield  [0]{\@secondoftwo}%
\providecommand \translation [1]{[#1]}%
\providecommand \BibitemOpen [0]{}%
\providecommand \bibitemStop [0]{}%
\providecommand \bibitemNoStop [0]{.\EOS\space}%
\providecommand \EOS [0]{\spacefactor3000\relax}%
\providecommand \BibitemShut  [1]{\csname bibitem#1\endcsname}%
\let\auto@bib@innerbib\@empty
\bibitem [{\citenamefont {Oliveira}\ \emph {et~al.}(2011)\citenamefont
  {Oliveira}, \citenamefont {Andrade},\ and\ \citenamefont
  {Herrmann}}]{Olivieira11}%
  \BibitemOpen
  \bibfield  {author} {\bibinfo {author} {\bibfnamefont {C.~L.~N.}\
  \bibnamefont {Oliveira}}, \bibinfo {author} {\bibfnamefont {J.~S.}\
  \bibnamefont {Andrade}}, \ and\ \bibinfo {author} {\bibfnamefont {H.~J.}\
  \bibnamefont {Herrmann}},\ }\href {\doibase 10.1103/PhysRevE.83.066307}
  {\bibfield  {journal} {\bibinfo  {journal} {Phys. Rev. E}\ }\textbf {\bibinfo
  {volume} {83}},\ \bibinfo {pages} {066307} (\bibinfo {year}
  {2011})}\BibitemShut {NoStop}%
\bibitem [{\citenamefont {Li}\ \emph {et~al.}(2015)\citenamefont {Li},
  \citenamefont {Huang},\ and\ \citenamefont {Faghri}}]{Li15}%
  \BibitemOpen
  \bibfield  {author} {\bibinfo {author} {\bibfnamefont {X.}~\bibnamefont
  {Li}}, \bibinfo {author} {\bibfnamefont {J.}~\bibnamefont {Huang}}, \ and\
  \bibinfo {author} {\bibfnamefont {A.}~\bibnamefont {Faghri}},\ }\href
  {\doibase http://dx.doi.org/10.1016/j.energy.2014.12.062} {\bibfield
  {journal} {\bibinfo  {journal} {Energy}\ }\textbf {\bibinfo {volume} {81}},\
  \bibinfo {pages} {489 } (\bibinfo {year} {2015})}\BibitemShut {NoStop}%
\bibitem [{\citenamefont {Penta}\ and\ \citenamefont
  {Ambrosi}(2015)}]{Penta15}%
  \BibitemOpen
  \bibfield  {author} {\bibinfo {author} {\bibfnamefont {R.}~\bibnamefont
  {Penta}}\ and\ \bibinfo {author} {\bibfnamefont {D.}~\bibnamefont
  {Ambrosi}},\ }\href@noop {} {\bibfield  {journal} {\bibinfo  {journal} {J.
  Theor. Biol.}\ }\textbf {\bibinfo {volume} {364}},\ \bibinfo {pages} {80}
  (\bibinfo {year} {2015})}\BibitemShut {NoStop}%
\bibitem [{\citenamefont {Kutsovsky}\ \emph {et~al.}(1996)\citenamefont
  {Kutsovsky}, \citenamefont {Scriven}, \citenamefont {Davis},\ and\
  \citenamefont {Hammer}}]{Kutsovsky96}%
  \BibitemOpen
  \bibfield  {author} {\bibinfo {author} {\bibfnamefont {Y.~E.}\ \bibnamefont
  {Kutsovsky}}, \bibinfo {author} {\bibfnamefont {L.~E.}\ \bibnamefont
  {Scriven}}, \bibinfo {author} {\bibfnamefont {H.~T.}\ \bibnamefont {Davis}},
  \ and\ \bibinfo {author} {\bibfnamefont {B.~E.}\ \bibnamefont {Hammer}},\
  }\href@noop {} {\bibfield  {journal} {\bibinfo  {journal} {Phys. Fluids}\
  }\textbf {\bibinfo {volume} {8}},\ \bibinfo {pages} {863} (\bibinfo {year}
  {1996})}\BibitemShut {NoStop}%
\bibitem [{\citenamefont {Mansfield}\ and\ \citenamefont
  {Issa}(1996)}]{Mansfield1996a}%
  \BibitemOpen
  \bibfield  {author} {\bibinfo {author} {\bibfnamefont {P.}~\bibnamefont
  {Mansfield}}\ and\ \bibinfo {author} {\bibfnamefont {B.}~\bibnamefont
  {Issa}},\ }\href {\doibase http://dx.doi.org/10.1006/jmra.1996.0189}
  {\bibfield  {journal} {\bibinfo  {journal} {J. Magn. Reson. Ser. A}\ }\textbf
  {\bibinfo {volume} {122}},\ \bibinfo {pages} {137} (\bibinfo {year}
  {1996})}\BibitemShut {NoStop}%
\bibitem [{\citenamefont {Datta}\ \emph {et~al.}(2013)\citenamefont {Datta},
  \citenamefont {Chiang}, \citenamefont {Ramakrishnan},\ and\ \citenamefont
  {Weitz}}]{Datta13}%
  \BibitemOpen
  \bibfield  {author} {\bibinfo {author} {\bibfnamefont {S.~S.}\ \bibnamefont
  {Datta}}, \bibinfo {author} {\bibfnamefont {H.}~\bibnamefont {Chiang}},
  \bibinfo {author} {\bibfnamefont {T.~S.}\ \bibnamefont {Ramakrishnan}}, \
  and\ \bibinfo {author} {\bibfnamefont {D.~A.}\ \bibnamefont {Weitz}},\
  }\href@noop {} {\bibfield  {journal} {\bibinfo  {journal} {Phys. Rev. Lett.}\
  }\textbf {\bibinfo {volume} {111}},\ \bibinfo {pages} {064501} (\bibinfo
  {year} {2013})}\BibitemShut {NoStop}%
\bibitem [{\citenamefont {Morad}\ and\ \citenamefont
  {Khalili}(2009)}]{Morad09}%
  \BibitemOpen
  \bibfield  {author} {\bibinfo {author} {\bibfnamefont {M.}~\bibnamefont
  {Morad}}\ and\ \bibinfo {author} {\bibfnamefont {A.}~\bibnamefont
  {Khalili}},\ }\href@noop {} {\bibfield  {journal} {\bibinfo  {journal} {Exp.
  Fluids}\ }\textbf {\bibinfo {volume} {46}},\ \bibinfo {pages} {323} (\bibinfo
  {year} {2009})}\BibitemShut {NoStop}%
\bibitem [{\citenamefont {Andrade}\ \emph {et~al.}(1997)\citenamefont
  {Andrade}, \citenamefont {Almeida}, \citenamefont {Mendes~Filho},
  \citenamefont {Havlin}, \citenamefont {Suki},\ and\ \citenamefont
  {Stanley}}]{Andrade97}%
  \BibitemOpen
  \bibfield  {author} {\bibinfo {author} {\bibfnamefont {J.~S.}\ \bibnamefont
  {Andrade}}, \bibinfo {author} {\bibfnamefont {M.~P.}\ \bibnamefont
  {Almeida}}, \bibinfo {author} {\bibfnamefont {J.}~\bibnamefont
  {Mendes~Filho}}, \bibinfo {author} {\bibfnamefont {S.}~\bibnamefont
  {Havlin}}, \bibinfo {author} {\bibfnamefont {B.}~\bibnamefont {Suki}}, \ and\
  \bibinfo {author} {\bibfnamefont {H.~E.}\ \bibnamefont {Stanley}},\ }\href
  {\doibase 10.1103/PhysRevLett.79.3901} {\bibfield  {journal} {\bibinfo
  {journal} {Phys. Rev. Lett.}\ }\textbf {\bibinfo {volume} {79}},\ \bibinfo
  {pages} {3901} (\bibinfo {year} {1997})}\BibitemShut {NoStop}%
\bibitem [{\citenamefont {Andrade}\ \emph {et~al.}(1999)\citenamefont
  {Andrade}, \citenamefont {Costa}, \citenamefont {Almeida}, \citenamefont
  {Makse},\ and\ \citenamefont {Stanley}}]{Andrade99}%
  \BibitemOpen
  \bibfield  {author} {\bibinfo {author} {\bibfnamefont {J.~S.}\ \bibnamefont
  {Andrade}}, \bibinfo {author} {\bibfnamefont {U.~M.~S.}\ \bibnamefont
  {Costa}}, \bibinfo {author} {\bibfnamefont {M.~P.}\ \bibnamefont {Almeida}},
  \bibinfo {author} {\bibfnamefont {H.~A.}\ \bibnamefont {Makse}}, \ and\
  \bibinfo {author} {\bibfnamefont {H.~E.}\ \bibnamefont {Stanley}},\ }\href
  {\doibase 10.1103/PhysRevLett.82.5249} {\bibfield  {journal} {\bibinfo
  {journal} {Phys. Rev. Lett.}\ }\textbf {\bibinfo {volume} {82}},\ \bibinfo
  {pages} {5249} (\bibinfo {year} {1999})}\BibitemShut {NoStop}%
\bibitem [{\citenamefont {Duda}\ \emph {et~al.}(2011)\citenamefont {Duda},
  \citenamefont {Koza},\ and\ \citenamefont {Matyka}}]{Duda11}%
  \BibitemOpen
  \bibfield  {author} {\bibinfo {author} {\bibfnamefont {A.}~\bibnamefont
  {Duda}}, \bibinfo {author} {\bibfnamefont {Z.}~\bibnamefont {Koza}}, \ and\
  \bibinfo {author} {\bibfnamefont {M.}~\bibnamefont {Matyka}},\ }\href@noop {}
  {\bibfield  {journal} {\bibinfo  {journal} {Phys. Rev. E}\ }\textbf {\bibinfo
  {volume} {84}},\ \bibinfo {pages} {036319} (\bibinfo {year}
  {2011})}\BibitemShut {NoStop}%
\bibitem [{\citenamefont {Bear}(1972)}]{Bear72}%
  \BibitemOpen
  \bibfield  {author} {\bibinfo {author} {\bibfnamefont {J.}~\bibnamefont
  {Bear}},\ }\href@noop {} {\emph {\bibinfo {title} {Dynamics of Fluids in
  Porous Media}}}\ (\bibinfo  {publisher} {Elsevier},\ \bibinfo {address} {New
  York},\ \bibinfo {year} {1972})\BibitemShut {NoStop}%
\bibitem [{\citenamefont {Bijeljic}\ \emph {et~al.}(2013)\citenamefont
  {Bijeljic}, \citenamefont {Raeini}, \citenamefont {Mostaghimi},\ and\
  \citenamefont {Blunt}}]{Bijeljic13}%
  \BibitemOpen
  \bibfield  {author} {\bibinfo {author} {\bibfnamefont {B.}~\bibnamefont
  {Bijeljic}}, \bibinfo {author} {\bibfnamefont {A.}~\bibnamefont {Raeini}},
  \bibinfo {author} {\bibfnamefont {P.}~\bibnamefont {Mostaghimi}}, \ and\
  \bibinfo {author} {\bibfnamefont {M.~J.}\ \bibnamefont {Blunt}},\ }\href@noop
  {} {\bibfield  {journal} {\bibinfo  {journal} {Phys. Rev. E}\ }\textbf
  {\bibinfo {volume} {87}},\ \bibinfo {pages} {013011} (\bibinfo {year}
  {2013})}\BibitemShut {NoStop}%
\bibitem [{\citenamefont {Ara\'ujo}\ \emph {et~al.}(2006)\citenamefont
  {Ara\'ujo}, \citenamefont {Bastos}, \citenamefont {{J. S. Andrade, Jr.}},\
  and\ \citenamefont {Herrmann}}]{Araujo06}%
  \BibitemOpen
  \bibfield  {author} {\bibinfo {author} {\bibfnamefont {A.~D.}\ \bibnamefont
  {Ara\'ujo}}, \bibinfo {author} {\bibfnamefont {W.~B.}\ \bibnamefont
  {Bastos}}, \bibinfo {author} {\bibnamefont {{J. S. Andrade, Jr.}}}, \ and\
  \bibinfo {author} {\bibfnamefont {H.~J.}\ \bibnamefont {Herrmann}},\ }\href
  {\doibase 0.1103/PhysRevE.74.010401} {\bibfield  {journal} {\bibinfo
  {journal} {Phys. Rev. E}\ }\textbf {\bibinfo {volume} {74}},\ \bibinfo
  {pages} {010401(R)} (\bibinfo {year} {2006})}\BibitemShut {NoStop}%
\bibitem [{\citenamefont {Lebon}\ \emph {et~al.}(1996)\citenamefont {Lebon},
  \citenamefont {Oger}, \citenamefont {Leblond}, \citenamefont {Hulin},
  \citenamefont {Martys},\ and\ \citenamefont {Schwartz}}]{Lebon96}%
  \BibitemOpen
  \bibfield  {author} {\bibinfo {author} {\bibfnamefont {L.}~\bibnamefont
  {Lebon}}, \bibinfo {author} {\bibfnamefont {L.}~\bibnamefont {Oger}},
  \bibinfo {author} {\bibfnamefont {J.}~\bibnamefont {Leblond}}, \bibinfo
  {author} {\bibfnamefont {J.~P.}\ \bibnamefont {Hulin}}, \bibinfo {author}
  {\bibfnamefont {N.~S.}\ \bibnamefont {Martys}}, \ and\ \bibinfo {author}
  {\bibfnamefont {L.~M.}\ \bibnamefont {Schwartz}},\ }\href {\doibase
  http://dx.doi.org/10.1063/1.868839} {\bibfield  {journal} {\bibinfo
  {journal} {Phys. Fluids}\ }\textbf {\bibinfo {volume} {8}},\ \bibinfo {pages}
  {293} (\bibinfo {year} {1996})}\BibitemShut {NoStop}%
\bibitem [{\citenamefont {Maier}\ \emph {et~al.}(1999)\citenamefont {Maier},
  \citenamefont {Kroll}, \citenamefont {Davis},\ and\ \citenamefont
  {Bernard}}]{Maier99}%
  \BibitemOpen
  \bibfield  {author} {\bibinfo {author} {\bibfnamefont {R.~S.}\ \bibnamefont
  {Maier}}, \bibinfo {author} {\bibfnamefont {D.~M.}\ \bibnamefont {Kroll}},
  \bibinfo {author} {\bibfnamefont {H.}~\bibnamefont {Davis}}, \ and\ \bibinfo
  {author} {\bibfnamefont {R.~S.}\ \bibnamefont {Bernard}},\ }\href@noop {}
  {\bibfield  {journal} {\bibinfo  {journal} {J. Colloid Interf. Sci.}\
  }\textbf {\bibinfo {volume} {217}},\ \bibinfo {pages} {341 } (\bibinfo {year}
  {1999})}\BibitemShut {NoStop}%
\bibitem [{\citenamefont {Maier}\ \emph {et~al.}(1998)\citenamefont {Maier},
  \citenamefont {Kroll}, \citenamefont {Kutsovsky}, \citenamefont {Davis},\
  and\ \citenamefont {Bernard}}]{Maier96}%
  \BibitemOpen
  \bibfield  {author} {\bibinfo {author} {\bibfnamefont {R.~S.}\ \bibnamefont
  {Maier}}, \bibinfo {author} {\bibfnamefont {D.~M.}\ \bibnamefont {Kroll}},
  \bibinfo {author} {\bibfnamefont {Y.~E.}\ \bibnamefont {Kutsovsky}}, \bibinfo
  {author} {\bibfnamefont {H.~T.}\ \bibnamefont {Davis}}, \ and\ \bibinfo
  {author} {\bibfnamefont {R.~S.}\ \bibnamefont {Bernard}},\ }\href@noop {}
  {\bibfield  {journal} {\bibinfo  {journal} {Phys. Fluids}\ }\textbf {\bibinfo
  {volume} {10}},\ \bibinfo {pages} {60} (\bibinfo {year} {1998})}\BibitemShut
  {NoStop}%
\bibitem [{\citenamefont {Siena}\ \emph {et~al.}(2014)\citenamefont {Siena},
  \citenamefont {Riva}, \citenamefont {Hyman}, \citenamefont {Winter},\ and\
  \citenamefont {Guadagnini}}]{Siena14}%
  \BibitemOpen
  \bibfield  {author} {\bibinfo {author} {\bibfnamefont {M.}~\bibnamefont
  {Siena}}, \bibinfo {author} {\bibfnamefont {M.}~\bibnamefont {Riva}},
  \bibinfo {author} {\bibfnamefont {J.~D.}\ \bibnamefont {Hyman}}, \bibinfo
  {author} {\bibfnamefont {C.~L.}\ \bibnamefont {Winter}}, \ and\ \bibinfo
  {author} {\bibfnamefont {A.}~\bibnamefont {Guadagnini}},\ }\href@noop {}
  {\bibfield  {journal} {\bibinfo  {journal} {Phys. Rev. E}\ }\textbf {\bibinfo
  {volume} {89}},\ \bibinfo {pages} {013018} (\bibinfo {year}
  {2014})}\BibitemShut {NoStop}%
\bibitem [{\citenamefont {Tomadakis}\ and\ \citenamefont
  {Sotirchos}(1993)}]{Tomadakis93}%
  \BibitemOpen
  \bibfield  {author} {\bibinfo {author} {\bibfnamefont {M.~M.}\ \bibnamefont
  {Tomadakis}}\ and\ \bibinfo {author} {\bibfnamefont {S.~V.}\ \bibnamefont
  {Sotirchos}},\ }\href@noop {} {\bibfield  {journal} {\bibinfo  {journal} {J.
  Chem. Phys.}\ }\textbf {\bibinfo {volume} {98}},\ \bibinfo {pages} {616}
  (\bibinfo {year} {1993})}\BibitemShut {NoStop}%
\bibitem [{\citenamefont {Succi}(2001)}]{Succi01}%
  \BibitemOpen
  \bibfield  {author} {\bibinfo {author} {\bibfnamefont {S.}~\bibnamefont
  {Succi}},\ }\href@noop {} {\emph {\bibinfo {title} {The Lattice Boltzmann
  Equation for Fluid Dynamics and Beyond}}}\ (\bibinfo  {publisher} {Clarendon
  Press},\ \bibinfo {address} {New York},\ \bibinfo {year} {2001})\BibitemShut
  {NoStop}%
\bibitem [{\citenamefont {Huber}\ \emph {et~al.}(2013)\citenamefont {Huber},
  \citenamefont {Parmigiani}, \citenamefont {Latt},\ and\ \citenamefont
  {Dufek}}]{Huber13}%
  \BibitemOpen
  \bibfield  {author} {\bibinfo {author} {\bibfnamefont {C.}~\bibnamefont
  {Huber}}, \bibinfo {author} {\bibfnamefont {A.}~\bibnamefont {Parmigiani}},
  \bibinfo {author} {\bibfnamefont {J.}~\bibnamefont {Latt}}, \ and\ \bibinfo
  {author} {\bibfnamefont {J.}~\bibnamefont {Dufek}},\ }\href@noop {}
  {\bibfield  {journal} {\bibinfo  {journal} {Water Resour. Res.}\ }\textbf
  {\bibinfo {volume} {49}},\ \bibinfo {pages} {6371} (\bibinfo {year}
  {2013})}\BibitemShut {NoStop}%
\bibitem [{\citenamefont {Pan}\ \emph {et~al.}(2006)\citenamefont {Pan},
  \citenamefont {Luo},\ and\ \citenamefont {Miller}}]{Pan06}%
  \BibitemOpen
  \bibfield  {author} {\bibinfo {author} {\bibfnamefont {C.}~\bibnamefont
  {Pan}}, \bibinfo {author} {\bibfnamefont {L.-S.}\ \bibnamefont {Luo}}, \ and\
  \bibinfo {author} {\bibfnamefont {C.~T.}\ \bibnamefont {Miller}},\
  }\href@noop {} {\bibfield  {journal} {\bibinfo  {journal} {Comput. Fluids}\
  }\textbf {\bibinfo {volume} {35}},\ \bibinfo {pages} {898 } (\bibinfo {year}
  {2006})}\BibitemShut {NoStop}%
\bibitem [{\citenamefont {Koza}\ \emph {et~al.}(2009)\citenamefont {Koza},
  \citenamefont {Matyka},\ and\ \citenamefont {Khalili}}]{Koza09}%
  \BibitemOpen
  \bibfield  {author} {\bibinfo {author} {\bibfnamefont {Z.}~\bibnamefont
  {Koza}}, \bibinfo {author} {\bibfnamefont {M.}~\bibnamefont {Matyka}}, \ and\
  \bibinfo {author} {\bibfnamefont {A.}~\bibnamefont {Khalili}},\ }\href
  {\doibase 10.1103/PhysRevE.79.066306} {\bibfield  {journal} {\bibinfo
  {journal} {Phys. Rev. E}\ }\textbf {\bibinfo {volume} {79}},\ \bibinfo
  {pages} {066306} (\bibinfo {year} {2009})}\BibitemShut {NoStop}%
\bibitem [{\citenamefont {Koza}\ \emph {et~al.}(2014)\citenamefont {Koza},
  \citenamefont {Kondrat},\ and\ \citenamefont {Suszczy\'nski}}]{Koza14}%
  \BibitemOpen
  \bibfield  {author} {\bibinfo {author} {\bibfnamefont {Z.}~\bibnamefont
  {Koza}}, \bibinfo {author} {\bibfnamefont {G.}~\bibnamefont {Kondrat}}, \
  and\ \bibinfo {author} {\bibfnamefont {K.}~\bibnamefont {Suszczy\'nski}},\
  }\href {http://stacks.iop.org/1742-5468/2014/i=11/a=P11005} {\bibfield
  {journal} {\bibinfo  {journal} {J. Stat. Mech.-Theory E.}\ }\textbf {\bibinfo
  {volume} {2014}},\ \bibinfo {pages} {P11005} (\bibinfo {year}
  {2014})}\BibitemShut {NoStop}%
\bibitem [{\citenamefont {Fleurant}\ and\ \citenamefont {van~der
  Lee}(2001)}]{Fleurant2001}%
  \BibitemOpen
  \bibfield  {author} {\bibinfo {author} {\bibfnamefont {C.}~\bibnamefont
  {Fleurant}}\ and\ \bibinfo {author} {\bibfnamefont {J.}~\bibnamefont {van~der
  Lee}},\ }\href {\doibase 10.1023/A:1011036929162} {\bibfield  {journal}
  {\bibinfo  {journal} {Math. Geol.}\ }\textbf {\bibinfo {volume} {33}},\
  \bibinfo {pages} {449} (\bibinfo {year} {2001})}\BibitemShut {NoStop}%
\bibitem [{\citenamefont {Sahimi}(1993)}]{Sahimi93}%
  \BibitemOpen
  \bibfield  {author} {\bibinfo {author} {\bibfnamefont {M.}~\bibnamefont
  {Sahimi}},\ }\href@noop {} {\bibfield  {journal} {\bibinfo  {journal} {Rev.
  Mod. Phys.}\ }\textbf {\bibinfo {volume} {65}},\ \bibinfo {pages} {1393}
  (\bibinfo {year} {1993})}\BibitemShut {NoStop}%
\bibitem [{\citenamefont {Hunt}(2009)}]{Hunt2009}%
  \BibitemOpen
  \bibfield  {author} {\bibinfo {author} {\bibfnamefont {A.}~\bibnamefont
  {Hunt}},\ }\href {http://www.springer.com/us/book/9783642100567} {\emph
  {\bibinfo {title} {{Percolation theory for flow in porous media}}}},\ Lecture
  Notes in Physics\ (\bibinfo  {publisher} {Springer},\ \bibinfo {address}
  {Berlin},\ \bibinfo {year} {2009})\BibitemShut {NoStop}%
\bibitem [{\citenamefont {Icardi}\ \emph {et~al.}(2014)\citenamefont {Icardi},
  \citenamefont {Boccardo}, \citenamefont {Marchisio}, \citenamefont {Tosco},\
  and\ \citenamefont {Sethi}}]{Icardi14}%
  \BibitemOpen
  \bibfield  {author} {\bibinfo {author} {\bibfnamefont {M.}~\bibnamefont
  {Icardi}}, \bibinfo {author} {\bibfnamefont {G.}~\bibnamefont {Boccardo}},
  \bibinfo {author} {\bibfnamefont {D.~L.}\ \bibnamefont {Marchisio}}, \bibinfo
  {author} {\bibfnamefont {T.}~\bibnamefont {Tosco}}, \ and\ \bibinfo {author}
  {\bibfnamefont {R.}~\bibnamefont {Sethi}},\ }\href@noop {} {\bibfield
  {journal} {\bibinfo  {journal} {Phys. Rev. E}\ }\textbf {\bibinfo {volume}
  {90}},\ \bibinfo {pages} {013032} (\bibinfo {year} {2014})}\BibitemShut
  {NoStop}%
\bibitem [{\citenamefont {Ghanbarian}\ \emph {et~al.}(2013)\citenamefont
  {Ghanbarian}, \citenamefont {Hunt}, \citenamefont {Ewing},\ and\
  \citenamefont {Sahimi}}]{Ghanbarian2013}%
  \BibitemOpen
  \bibfield  {author} {\bibinfo {author} {\bibfnamefont {B.}~\bibnamefont
  {Ghanbarian}}, \bibinfo {author} {\bibfnamefont {A.~G.}\ \bibnamefont
  {Hunt}}, \bibinfo {author} {\bibfnamefont {R.~P.}\ \bibnamefont {Ewing}}, \
  and\ \bibinfo {author} {\bibfnamefont {M.}~\bibnamefont {Sahimi}},\ }\href
  {http://dx.doi.org/10.2136/sssaj2012.0435} {\bibfield  {journal} {\bibinfo
  {journal} {Soil Sci. Soc. Am. J.}\ }\textbf {\bibinfo {volume} {77}},\
  \bibinfo {pages} {1461} (\bibinfo {year} {2013})}\BibitemShut {NoStop}%
\bibitem [{\citenamefont {Rong}\ \emph {et~al.}(2013)\citenamefont {Rong},
  \citenamefont {Dong},\ and\ \citenamefont {Yu}}]{Rong13}%
  \BibitemOpen
  \bibfield  {author} {\bibinfo {author} {\bibfnamefont {L.}~\bibnamefont
  {Rong}}, \bibinfo {author} {\bibfnamefont {K.}~\bibnamefont {Dong}}, \ and\
  \bibinfo {author} {\bibfnamefont {A.}~\bibnamefont {Yu}},\ }\href@noop {}
  {\bibfield  {journal} {\bibinfo  {journal} {Chem. Eng. Sci.}\ }\textbf
  {\bibinfo {volume} {99}},\ \bibinfo {pages} {44 } (\bibinfo {year}
  {2013})}\BibitemShut {NoStop}%
\end{thebibliography}%

\end{document}